\documentclass[%
reprint,
groupedaddress,
 amsmath,amssymb,
 aps,
]{revtex4-1}

\usepackage{comment}
\usepackage{graphicx}
\usepackage{dcolumn}
\usepackage{bm}

\begin{document}

\title{Morphological changes of amphiphilic molecular assemblies 
induced by chemical reaction}

\author{Koh M. Nakagawa}%
\author{Hiroshi Noguchi}%
\email{noguchi@issp.u-tokyo.ac.jp}
\affiliation{%
 Institute for Solid State Physics, University of Tokyo, Kashiwa, Chiba 277-8581, Japan\\
}%

\date{\today}

\begin{abstract}
Shape transformations of amphiphilic molecular assemblies induced by chemical reaction 
are studied using coarse-grained molecular simulations.
A binding reaction between hydrophilic and hydrophobic molecules is considered.
It is found that the reaction induces 
transformation of an oil droplet to a tubular vesicle via 
 bicelles and vesicles with discoidal arms.
The discoidal arms close into vesicles, which are subsequently
fused into the tubular vesicle.
Under the chemical reaction,
the bicelle-to-vesicle transition occurs at smaller sizes than 
in the absence of the hydrophobic molecules.
It is revealed that the enhancement of this transition is
due to embedded hydrophobic particles that
reduce the membrane bending rigidity.
\end{abstract}

\pacs{Valid PACS appear here}
\maketitle

\section{\label{Intro}Introduction}
Amphiphilic molecules in an aqueous solution spontaneously aggregate into
various structures \cite{hamley2008introduction}. Depending on the
molecular structure, different 
self-assembled structures appear, {\it e.g.}, spherical and worm-like micelles, vesicles, bicontinuous structures, etc.
In particular, the morphology of vesicles attracts the attention of many
researchers, as this is the simplest model of a cell membrane. 
Bilayer vesicles exhibit various shapes,
{\it e.g.}, discocyte (red-blood-cell shape),
single and double stomatocytes, tubulation, pearl-necklace shapes, and so on \cite{lipo95,Seifert1997a,svet09,yana08,saka14}. 
Thus, static vesicle shapes have been well explored.
In living cells, however, membrane sizes are dynamically changed by the moments
of the fusion and fission of small vesicles.
The lipid composition of biomembranes varies depending on the organelle and cell type, and
the asymmetric distribution of the lipids between inner and outer leaflets
can yield a spontaneous membrane curvature.
These compositions are controlled by metabolic reactions on Golgi apparatus and 
lipid droplets \cite{fuji11,pol14}, and by vesicular transport in cells.
However, the evolution dynamics of lipid droplets on endoplasmic reticulum and the control of their sizes and shapes
are not understood so far.

Recently, the shape evolution of vesicles and micelles induced by chemical
reactions has been experimentally observed \cite{Suzuki2009,Toyota2006,Takakura2003}.
In experiments conducted by Toyota et al. \cite{Toyota2006}, amphiphilic 
molecules were divided into hydrophobic and hydrophilic
molecules by hydrolysis. It was found that the resultant hydrophobic molecules 
induce a decrease in the spontaneous curvature of amphiphilic molecular 
assemblies, leading to various shape changes.
Initially, these amphiphilic molecules aggregate into a spherical micelle
structure. As the chemical reaction progresses, this subsequently
transforms into a tubular micelle, spherical vesicle, tubular vesicle, 
stomatocyte, and nested vesicle. Finally,
the amphiphilic molecules form oil droplets.
Through this reaction, a nanometer order structure (micelles) surprisingly
changes to micrometer order structure (vesicles and droplets). 
This chemical reaction progresses slowly compared to vesicle shape
deformation.

In experiments conducted by Takakura et al. \cite{Takakura2003}, 
hydrophobic and hydrophilic molecules are bound into amphiphilic molecules by
dehydro-condensation. 
Initially, hydrophobic molecules are formed in an oil
droplet in water. The chemical reaction
occurs on the droplet surface and decreases the surface tension. Consequently,
various complex structures ({\it e.g.}, multiple vesicles
connected together by narrow necks) are observed.
Using this chemical reaction, Takakura produced a self-reproducing system,
in which new vesicles (daughter vesicles) are produced 
from oil droplets in the vesicles (mother vesicles). 
Self-reproduction
 is one of the main functions of protocells \cite{zepi08,stan10,kuri11}, and understanding
 the detailed mechanism of shape transformation induced by chemical reactions
is important to improving self-reproduction control.

Although various shape deformations induced by chemical reactions are observed {\it in vitro} and {\it in vivo},
 their mechanisms are not understood so far.
The aim of this study is to clarify the shape transformation of the binding reaction to form amphiphilic molecules
using coarse-grained molecular simulations.
The lengths and time scales of all-atom simulations are still limited and, in
order to simulate large-scale dynamics,
various coarse-grained molecules are developed \cite{Venturoli2006,marr09,nogu09}.
In comparison with thermal equilibrium states,
 non-equilibrium dynamics including chemical reactions have been explored to a much lesser extent.
Here, we propose a simple binding reaction model between one hydrophilic and one hydrophobic particle
to form an amphiphilic molecule.
We show how this reaction modifies the membrane properties and induces morphological changes.

In Section \ref{chp1}, our
simulation model is described while, in Section \ref{chp2}, the shape transformation
induced by chemical reaction is explained. 
Section \ref{chp3} presents the effects of hydrophobic
particles embedded in the bilayer membrane and
Section \ref{sum} contains the summary and discussion.

\section{\label{chp1}Simulation Model and Method}
\subsection{\label{chp1:sub1}Dissipative Particle Dynamics}

Dissipative Particle Dynamics (DPD) is a coarse-grained molecular simulation method to take into account hydrodynamic interactions
\cite{hoog92,Espanol1995,Groot1997}. 
The DPD method is applied to complex
fluids such as amphiphilic molecules \cite{Venturoli2006,Venturoli1999,groo01,Yamamoto2002,gao08,rama09,deme10,thak11,arai13,wu13,fedo14} and polymers \cite{Groot1997,Tschop1998,niku07,Li2008}. 
In this method,
several atoms are coarse-grained into one DPD particle, so long-term
simulation can be performed compared to atomic-scale molecular dynamics. 

In the DPD method, time evolutions are described using Newton's equation of
motion with a pairwise Langevin thermostat, such that
\begin{align}
  \label{eq:dpdeq}
 \frac{d\bm{r}_{i}}{dt}&=\bm{v}_{i}, \ \  m\frac{d\bm{v}_{i}}{dt} = \bm{f}_{i}, \\ \nonumber
 \bm{f}_{i}&=\Sigma_{j \neq i}(\bm{F}_{ij}^{\mathrm C}+\bm{F}_{ij}^{\mathrm D}+\bm{F}_{ij}^{\mathrm R}).
\end{align}
The conservative force, $\bm{F}_{ij}^{\mathrm C}$, is given by
\begin{equation}
 \bm{F}_{ij}^{\mathrm C} = a_{ij} (1 - r_{ij}/\sigma)\Theta(1 - r_{ij}/\sigma)\hat{\bm{r}}_{ij},
\end{equation}
where $a_{ij}$ is the maximum repulsive force strength, $\sigma$ is the cut-off length, 
$\bm{r}_{ij}=\bm{r}_{i} - \bm{r}_{j}$, and $\hat{\bm{r}}_{ij}=\bm{r}_{ij} / |\bm{r}_{ij}|$.
The unit step function, $\Theta(x)$, is defined
as 
\begin{equation}
 \Theta(x) = \begin{cases}
	      1 & (x < 0), \\
	      0 & (\mathrm{otherwise}).
	     \end{cases}
\end{equation}
The dissipation force, $\bm{F}_{ij}^{\mathrm D}$, is given by
\begin{equation}
 \bm{F}_{ij}^{\mathrm D} = -\gamma_{ij}\omega^{\mathrm D}(r_{ij})(\hat{\bm{r}}_{ij}\cdot \bm{v}_{ij})\Theta(1 - r_{ij}/\sigma)\hat{\bm{r}}_{ij}, 
\end{equation}
where $\bm{v}_{ij} = \bm{v}_{i} - \bm{v}_{j}$, $\gamma_{ij}$ is a friction
parameter and $\omega^{\mathrm D}(r)$ is a weight function.
The random force, $\bm{F}_{ij}^{\mathrm R}$, is given by
\begin{equation}
 \bm{F}_{ij}^{\mathrm R} =  \sigma_{ij}\omega^{\mathrm R}(r_{ij})\zeta_{ij}(t)\Theta(1 - r_{ij}/\sigma)\hat{\bm{r}}_{ij},
\end{equation}
where $\sigma_{ij}$ is noise strength and $\zeta_{ij}(t)$ is a Gaussian random number with the following properties:
\begin{align}
 \langle\zeta_{ij}(t) \rangle &= 0, \\
 \langle\zeta_{ij}(t)\zeta_{kl}(t')\rangle &= 2(\delta_{ik}\delta_{jl} + \delta_{il}\delta_{jk})\delta(t-t').
\end{align}
To satisfy the fluctuation dissipation theorem, $\gamma_{ij}$,
$\sigma_{ij}$, $\omega^{\mathrm D}(r)$, and $\omega^{\mathrm R}(r)$ should be related via
\cite{Groot1997,Espanol1995}
\begin{equation}
  \sigma_{ij}^{2}[\omega^{\mathrm R}(r)]^{2}= \gamma_{ij} k_{\mathrm{B}}T\omega^{\mathrm D}(r).
\end{equation}
We choose $\omega^{\mathrm D}(r)=[\omega^{\mathrm R}(r)]^{2}=(1-r/\sigma )^{2}$.

In the DPD method, the random and dissipation forces act as pairwise
forces. Since the action-reaction law is satisfied for each
particle pair, the momentum conservation 
law is satisfied. Thus, the hydrodynamic interactions are taken into account.

\subsection{\label{chp1:sub2}Molecular Model}
In this paper, W, H, and T denote a coarse-grained
water, hydrophilic, and hydrophobic particle, respectively. For simplicity, the
amphiphilic molecule is modeled by two soft particles connected by a spring force, such that
\begin{equation}\label{eq:bond}
 \bm{F}_{ij}^{\mathrm B} = C(1-r_{ij}/r_{\mathrm b})\hat{\bm{r}}_{ij},
\end{equation}
where $C$ is the bond strength and $r_{\mathrm b}$ is the bond length. 
We set $C=480k_{\mathrm B}T/\sigma$ and $r_{\mathrm b}=\sigma$.
In order to form a bilayer membrane and an oil droplet,
 $a_{ij}$ and $\gamma_{ij}$ are set as shown in Table \ref{tb:intr}.

\begin{table}[h]
\small
\caption{Interaction parameters $a_{ij}$ and $\gamma_{ij}$ with units
 $k_{\mathrm B}T/\sigma$ and $\sqrt{mk_{\mathrm B}T}/\sigma$, respectively.
}
 \begin{tabular*}{0.5\textwidth}{@{\extracolsep{\fill}}lll}
  \hline
  Bead pair  & $a_{ij}$  &$\gamma_{ij}$  \\ 
  \hline
  WW          & 100 & 72   \\ 
  HH          & 100 & 72   \\
  TT          & 100 & 72   \\
  WH          & 100 & 72   \\
  WT          & 300 & 288   \\
  HT          & 200 & 144   \\
  \hline
 \end{tabular*}
\label{tb:intr}
\end{table}

The equations of motion are discretized by the modified Verlet algorithm \cite{Groot1997}, such that
\begin{align}
 \bm{r}_{i}(t+\Delta t) &= \bm{r}_{i}(t) + \bm{v}_{i}(t)\Delta t + \frac{\Delta t^2}{2m}\bm{f}_{i}(t), \\
 \tilde{\bm{v}}_{i}(t+\Delta t) &= \bm{v}_{i}(t) + \frac{\lambda\Delta t}{m}\bm{f}_{i}(t), \\
 \bm{f}_{i}(t+\Delta t) &= \bm{f}_{i}(\bm{r}_{i}(t+\Delta t),\tilde{\bm{v}}_{i}(t+\Delta t)), \\
 \bm{v}_{i}(t+\Delta t) &= \bm{v}_{i}(t) + \frac{\Delta t}{2m} (\bm{f}_{i}(t) + \bm{f}_{i}(t+\Delta t)).
\end{align}
Note that a large time step yields artifacts
in DPD simulations \cite{Allen2006}. To avoid the emergence of such
artifacts, the time step, $\Delta t$, is set to $0.005\sigma\sqrt{m/k_{\mathrm B}T}$.
The factor, $\lambda$, is often set to 0.65 in DPD simulations \cite{Groot1997}. 
We compared the temperature difference between $\lambda=0.5$ and $\lambda=0.65$,
 but no apparent difference was detected between these values for $\Delta t = 0.005\sigma\sqrt{m/k_{\mathrm B}T}$.
In both settings, the deviations of the configurational and kinetic temperatures from the specified value are less than 0.5\%. 
Thus, $\lambda$ is set to 0.5 in this study.
The DPD program is originally implemented using CUDA \cite{nguyen2007gpu} .

All simulations are performed in the $NVT$ ensemble (constant particle 
number, $N$, volume, $V$, and temperature, $T$) with the periodic boundary condition. 
The total number of particles is $N=N_{\mathrm H}+N_{\mathrm T}+N_{\mathrm W}$,
while the number density $\rho$ of all particles is fixed as $\rho=3\sigma^{-3}$. 
We use reduced units with $\sigma$ as the unit of length, $k_{\mathrm{B}}T$ as the unit of energy, and $\tau$ as the unit of time, 
where $\tau = \sigma^{2}/D$ with the diffusion constant, $D = 0.082\sigma\sqrt{k_{\mathrm B}T/m}$, of W particles. 
The viscosity, $\eta$, of the W particle fluids is obtained as $\eta {\sigma}^{2}/\sqrt{mk_{\mathrm B}T}= 4.0 \pm 0.1$
from the stress calculation in simple shear flow \cite{nogu07a}.
Thus, our simulation condition yields a high Schmidt number for the DPD particles: $Sc=\eta/m\rho D= 16$.
Since the momentum propagates significantly more rapidly than the mass,
hydrodynamic interactions are taken into account in the diffusion of the DPD particles.
As the typical lengths of lipids and other amphiphilic molecules is $1$--$3$ nm, we can consider to be $\sigma \sim 1$ nm.
Thus, the simulation time unit is estimated as $\tau \sim 0.6$ ns in water ($\eta=0.8$ mPa$\cdot$s) at room temperature.

\subsection{Membrane Properties in the Absence of Hydrophobic Molecules}

\subsubsection{\label{chp1:sub3:sub1}Bending Rigidity.~~}

The amphiphilic molecules, consisting of one hydrophilic (H) and one hydrophobic (T) particle,
self-assemble into a fluid bilayer membrane in water.
We estimate the bending rigidity, $\kappa$, of the membrane from the thermal fluctuations of a flat
membrane \cite{Venturoli2006,goet99,shib11}.
Using Monge parameterization, the membrane shape is described by 
the membrane height, $h(x,y)$, above the $xy$ plane.
According to the equipartition theorem, the thermal fluctuations of the membrane and $\kappa$ are related via 
the equation \cite{safran1994statistical}
\begin{equation}
 \langle |h(\bm{q})|^{2} \rangle = \frac{k_{\mathrm B}T}{\gamma |\bm{q}|^{2} + \kappa |\bm{q}|^{4}},
  \label{eq:bend}
\end{equation}
where $\gamma$ is the surface tension of the membrane.
The Fourier mode, $h(\bm{q})$, of $h(x,y)$ is given by
\begin{eqnarray}
 h(\bm{q}) = \frac{1}{\sqrt{A}}\int h(x,y) \exp (-i\bm{q}\cdot\bm{r}) dxdy,
\end{eqnarray}
where $A$ is the membrane area.
We estimate $\kappa$ from eqn~(\ref{eq:bend}) for a tensionless membrane ($\gamma = 0$). 
The size of the simulation box is set to $32\sigma \times32\sigma \times32\sigma$. 
For the tensionless membrane, the number of amphiphiles, $N_{\mathrm{amp}}$, is 4700
and the area per amphiphile is $0.4357\sigma^2$. The bending rigidity is obtained as $\kappa=18.0 \pm 0.3 k_{\mathrm{B}}T$.

\subsubsection{\label{chp1:sub3:sub2}Edge Line Tension.~~}
We estimate the line tension, $\Gamma$, of the membrane edge 
from a membrane strip.
The line tension is calculated from \cite{Tolpekina2004,nogu11}
\begin{equation}
 \Gamma =  \langle \frac{P_{xx} + P_{zz}}{2} - P_{yy} \rangle  \frac{L_{z}L_{x}}{2},
\end{equation}
with
\begin{equation}
 P_{\alpha\alpha} = \frac{1}{V}\Big(Nk_{\mathrm B}T + \sum_{i>j}(\alpha_{i} - \alpha_{j}) F^{\mathrm C}_{ij\alpha}\Big),
\end{equation}
where $\alpha \in x,y,z$, $F^{\mathrm C}_{ij\alpha}$ is the $\alpha$ component of $\bm{F}_{ij}^{\mathrm C}$,
 and $L_{\alpha}$ is the simulation box length along the $\alpha$ axis. 
The strip is along the $y$ axis and
the edge line length is $2L_{y}$.
We estimate $\Gamma$ for several line lengths, $L_{y}/\sigma$, from 17 to 21, at $L_{z}=L_{x}=32\sigma$ and $N_{\mathrm{amp}}=1000$,
and take the average for different values of $L_{y}$.
The line tension is obtained as $\Gamma=6.1 \pm 0.2 k_{\mathrm{B}}T/\sigma$.

\subsection{\label{chp1:sub3}Chemical Reaction Model}
In this paper, we consider only a binding chemical reaction between hydrophilic and hydrophobic molecules forming an amphiphilic molecule, 
so that the inverse reaction (bond dissociation) is neglected.
The chemical reaction is phenomenologically introduced as a stochastic process for a neighbor pair of H and T particles.
For the reacted pair,  the spring force, eqn (\ref{eq:bond}), is added between the H and T particles. 
In each simulation time step, 
the binding probability, $P(r_{ij})$, is calculated from the particle distance, with
\begin{equation}
 P(r_{ij})\Delta t =
  \begin{cases}
   p_{0}\Theta(1-r_{ij}/\sigma)\exp \left[-\dfrac{r_{ij}^{2}}{2x^{2}_{0}}\right] &((i,j) = (H,T) \: \\ & \qquad \mbox{or}  \: (T,H)), \\
   0 & (\mathrm{otherwise}),
  \end{cases}
\end{equation}
where $p_{0}$ is the chemical reaction rate and $x_{0}$ is the chemical reaction length scale.
We set $p_{0}=1$ and $\sigma^{2}/x_{0}^2=20$. Thus, $x_{0} \simeq 0.223\sigma$.

\section{\label{chp2}Shape Deformation Induced by Chemical Reaction}
\begin{figure}
    \centering
     \includegraphics[width=70mm]{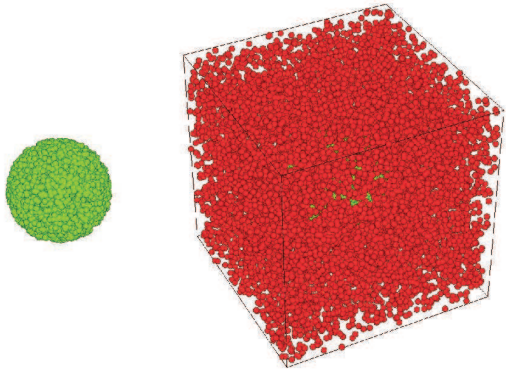}

    \caption{Snapshot of initial state of an oil
    droplet at $N_{\mathrm H}=12403$, $N_\mathrm {T}=6566$, and $N=145920$.
    The hydrophobic (green) particles are only shown in the left panel.
    Hydrophilic (red) particles are also shown in the right panel.
    Water particles are not shown for clarity. }
    \label{fig:chp2one}
   \end{figure}

As an initial condition, hydrophobic particles are centered in water, as shown in Fig. \ref{fig:chp2one}. 
The hydrophobic particles form a spherical droplet through hydrophobic interactions.
This droplet structure is called an oil droplet in the
following. 
The hydrophilic particles are dispersed in water
and no amphiphilic molecules initially exist. 
The number of hydrophilic, hydrophobic and total particles are $N_{\mathrm H}=12403$,
 $N_\mathrm {T}=6566$, and $N=145920$, respectively. 
After equilibrating the system for $4\tau$ steps,
the chemical reaction as described in Section \ref{chp1:sub3} is activated.

\begin{figure}
    \centering
     \includegraphics[width=85mm]{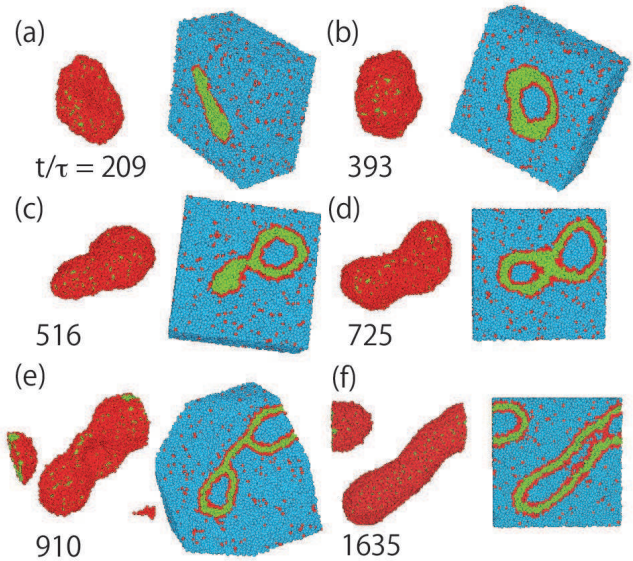}

    \caption{Sequential snapshots of an amphiphilic molecular assembly under the binding reaction at $N_{\mathrm H}=12403$,
 $N_\mathrm {T}=6566$, and $N=145920$. 
    The number represent the simulation time $t/\tau$.
    Sliced snapshots with water (blue) particles are shown on the right side.}
    \label{fig:chp2two}
   \end{figure}

A typical example of shape deformation dynamics is shown in Fig. \ref{fig:chp2two}.
The binding reaction occurs on the surface of an oil
droplet. 
Through this reaction,
 the surface tension of the oil droplet decreases so that
 the area increases. As the chemical reaction progresses, the oil droplet
transforms to a disk micelle called bicelle (see Fig. \ref{fig:chp2two}(a)). 
Subsequently, the bicelle
closes into a vesicle  via a bowl-like shape (see Fig. \ref{fig:chp2two}(b)).
This closing transition will be discussed later, in Section \ref{chp3}. 
In this spherical vesicle, 
unreacted hydrophobic
particles are still embedded in the bilayer membrane.
As the chemical reaction proceeds,
a discoidal arm grows from the vesicle (see Fig. \ref{fig:chp2two}(c)), which then
closes to form an additional vesicle
(see Fig. \ref{fig:chp2two}(d)). 
The two vesicles are connected to a stalk-like structure, which is observed during
the membrane fusion \cite{Noguchi2001a,mull11,riss14}.
The unreacted hydrophobic particles are concentrated on the stalk region, and the
two vesicles are fused into a single large tubular vesicle (see Fig. \ref{fig:chp2two}(f)). 
During this fusion process, the region connecting two vesicles
expands into a flat membrane, as shown in
Fig. \ref{fig:chp2two}(e). 
This fusion pathway is similar to the stalk-bending (or stalk-leaky) pathway
reported in Ref. \cite{Noguchi2001a,mull11,riss14}.
In the present system, 
 the stalk expansion is provoked by the chemical reaction.

\begin{figure}
    \centering
     \includegraphics[width=70mm]{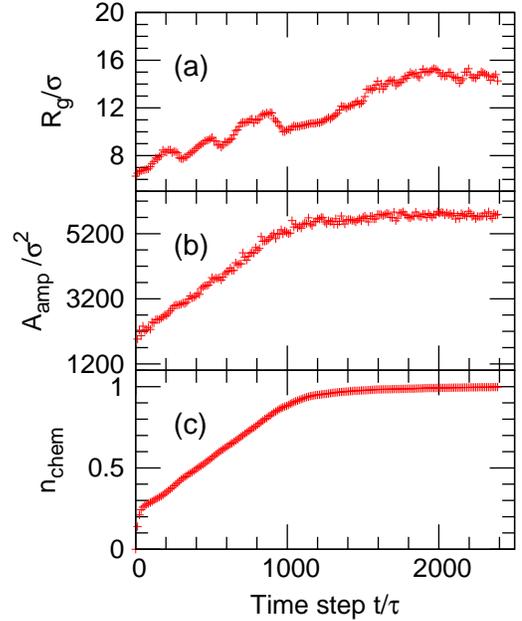}
    \caption{Time development of (a) the radius of gyration, $R_{\mathrm g}$,
    (b) area, $A_{\mathrm{amp}}$, of the amphiphilic molecular assembly, and 
    (c) chemical reaction ratio, $n_{\mathrm{chem}}=N_{\mathrm {amp}}/N_{\mathrm T}$. 
    The same data as in Fig. \ref{fig:chp2two} is shown. }
    \label{fig:chp2thr}
   \end{figure}

To quantify the shape deformation induced by the chemical reaction,
we calculate three quantities: the radius of gyration, $R_{\mathrm g}$, 
the surface area, $A_{\mathrm{amp}}$, of the amphiphilic molecular assembly, and 
the ratio of chemically reacted particles, $n_{\mathrm{chem}}=N_{\mathrm{amp}}/N_{\mathrm T}$,
as shown in Fig. \ref{fig:chp2thr}.
The radius of gyration is defined as 
$ R_{\mathrm g}^{2} = \sum_{i} \left| \bm{r}_{i} - \bm{r}_{\mathrm G} \right|^{2}/N_{\mathrm{ass}} ,$
where $\bm{r}_{\mathrm G}$ and $N_{\mathrm{ass}}$ are the center of mass and number of particles in the amphiphilic molecular assembly, respectively.
At $t/\tau\lesssim 1000$, $R_{\mathrm g}$ increases and decreases repeatedly, as shown in
Fig. \ref{fig:chp2thr}(a). 
These increases and decreases indicate
discoidal arm growth and vesicle closure, respectively.

$A_{\mathrm{amp}}$ is calculated as follows. 
First, the simulation box is divided
 into small cubic boxes with side length $\sigma$, 
and then the number, $n_{\mathrm{amp}}$, 
of hydrophobic and reacted hydrophilic particles
 in each box is counted. Second,  the boxes in which  $n_{\mathrm{amp}} \geq n_{\mathrm{cut}}$ are extracted.
Here, the cut-off number, $n_{\mathrm{cut}}$, is set to 2. 
Third, the number, $N_{\mathrm {wface}}$,
of the small box faces that are exposed to water is counted ($n_{\mathrm{amp}} < n_{\mathrm{cut}}$).
The surface area is estimated from $A_{\mathrm{amp}}=N_{\mathrm {wface}}\sigma^2$.

$A_{\mathrm{amp}}$ monotonically increases, as shown in Fig. \ref{fig:chp2thr}(b).
Since the amphiphilic particles 
prefer to be on the surface of the amphiphilic
assembly, 
the increase in $A_{\mathrm{amp}}$ is linear to an increase in the reaction ratio, $n_{\mathrm{chem}}$. 
This fact is clear from comparison between Fig. \ref{fig:chp2thr} (b) and Fig. \ref{fig:chp2thr} (c).

\begin{figure}[thbp]
    \centering
     \includegraphics[width=70mm]{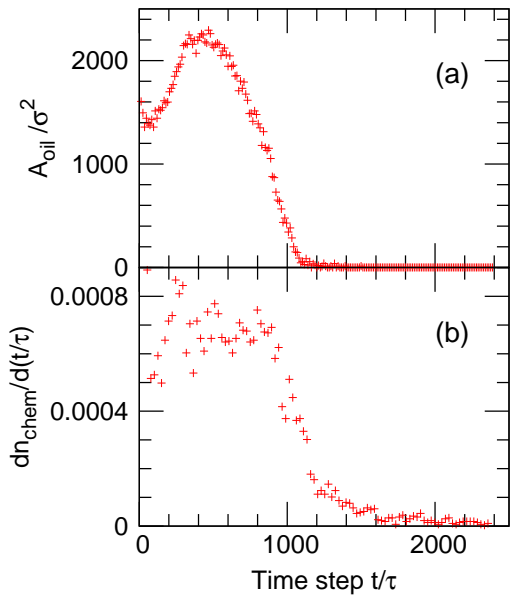}
    \caption{Time development of (a) the area $A_{\mathrm{oil}}$ of unreacted hydrophobic particles 
     and (b) the chemical reaction rate $dn_{\mathrm{chem}}/dt$. The same data in Fig. \ref{fig:chp2two} is shown.  }
    \label{fig:chp2sev}
   \end{figure}

To further clarify the chemical
reaction progression, we compare the surface
areas, $A_{\mathrm{oil}}$, of the unreacted hydrophobic particles and 
the binding reaction rate, $dn_{\mathrm{chem}}/dt$ (see Fig. \ref{fig:chp2sev}).
$A_{\mathrm{oil}}$ is determined using the same procedure as for
 $A_{\mathrm{amp}}$.
As shown in Figs. \ref{fig:chp2sev} (a) and (b), there is a positive
correlation between the two quantities. Under the fluctuation of 
the amphiphilic assembly, 
the chemical reaction occurs on the surface of  the oil droplet.
Thus, $dn_{\mathrm{chem}}/dt$ is roughly proportional to $A_{\mathrm{oil}}$. 

\begin{figure}
    \centering
     \includegraphics[width=85mm]{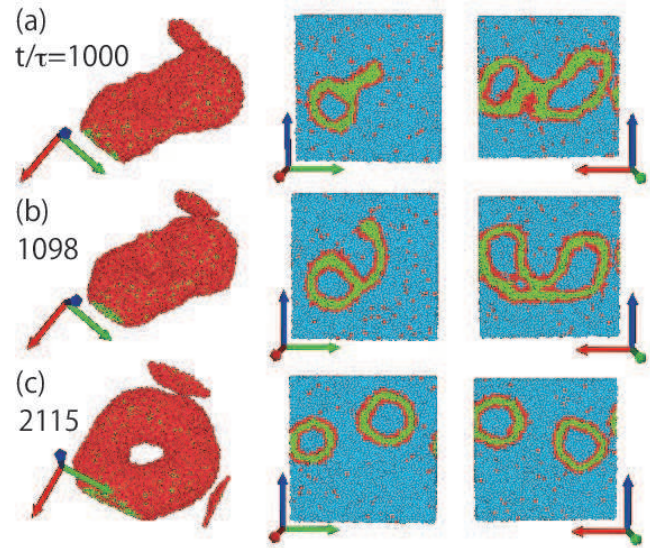}
    \caption{Sequential snapshots of a shape transformation from an oil droplet into a toroidal vesicle at 
$N_{\mathrm H}=13132$, $N_{\mathrm T}=10943$, and $N=145920$. 
             Sliced views are shown in the middle and right rows and two sliced sections are perpendicular to each other. 
             The red, green, and blue arrows indicate the $x$, $y$, and $z$ directions, respectively.
             }
    \label{fig:chp2six}
   \end{figure}

For a larger oil droplet, more vesicles and arms are
produced as intermediate states under the chemical reaction.
We investigate the shape development at four parameter sets, as summarized in Table \ref{tb:chp2:one}.
Four independent runs are performed for each condition.
In cases of $N_{\mathrm H}=13132$, $N_{\mathrm T}=10943$, and $N=145920$
(bottom row in Table \ref{tb:chp2:one}), 
three vesicles are produced from one oil droplet. 
In this case, an alternative final shape, that is, a toroidal shape, can be formed as, shown in Fig. \ref{fig:chp2six}.
Whether the final shape is a tubular or toroidal vesicle 
depends on the existence of a certain intermediate shape, as shown in
Fig. \ref{fig:chp2six} (a).
This intermediate shape is composed of a tubular vesicle and
a flat membrane, which is attached to the side of the tubular vesicle.
The tubular vesicle is bent around
the disk. Then, the ends of the tubular
vesicle are fused together, 
resulting in the formation of the toroidal vesicle
(see Figs. \ref{fig:chp2six} (b) and (c)).
This bending process is similar to the  bending of the stalk 
 shown in Fig. \ref{fig:chp2two}(e).
Thus, the bending around the neighboring compartment is a type of generic dynamics for amphiphilic systems.

\begin{table}[h]
 \small
 \caption{Final shapes generated from oil droplets for four
 simulation settings. Four independent runs are executed for each setting. The fourth/fifth row indicate the number of simulation samples with tubular or toroidal vesicle final shapes.}
 \begin{tabular*}{0.5\textwidth}{@{\extracolsep{\fill}}lllll}
  \hline
  $N_{\mathrm H}$ & $N_{\mathrm T}$ & $N$ &  Tubular & Toroidal  \\ 
  \hline
  12403   & 6566    & 145920  &  4  & 0 \\
  11424   & 8601    & 134400  &  3  & 1 \\
  13132   & 10214   & 145920  &  4  & 0 \\
  13132   & 10943   & 145920  &  3  & 1 \\
  \hline
 \end{tabular*}
\label{tb:chp2:one}
\end{table}

\section{\label{chp3}Effects of Embedded Hydrophobic Particles}

In the simulations described in Section \ref{chp2}, we notice that
the shape transition from bicelle to vesicle is enhanced by the chemical
reaction. The vesicle is formed from smaller membranes in comparison 
to those only consisting of amphiphilic molecules.
Since the transition occurs under non-equilibrium conditions,
this enhancement may be caused by some dynamic effects due to inhomogeneity of the reaction
and hydrodynamic interactions, or by changes in the membrane properties due to embedded hydrophobic particles.
To clarify whether the enhancement is caused by dynamic or static effects,
we simulate the bicelle-to-vesicle transition of the membrane with  embedded hydrophobic particles
in the absence of the chemical reaction (Section \ref{chp3:sub3}).
It is revealed that the transition is enhanced by the embedded particles.
Next, we investigate the dependence of the transition on the membrane properties (Section \ref{chp3:sub1} and \ref{chp3:sub2})
and conclude that the enhancement is caused by the reduction of the bending rigidity (Section \ref{chp3:sub4}).

\subsection{\label{chp3:sub3}Enhancement of Bicelle-to-Vesicle Transition}

We use the method proposed by Hu et al. \cite{Hu2012} to calculate the free
energy barrier between bicelle and vesicle. 
This method was originally used for determining the Gaussian curvature modulus, $\bar{\kappa}$.
Here, we briefly summarize this method.

For a homogeneous membrane, the shape transition from bicelle to vesicle is understood by
the competition between the edge-line and bending energies of the membrane \cite{Fromherz1983}.
The bicelle has edge energy of $2\pi R_{\mathrm {dis}}\Gamma$ and no bending energy,
where $R_{\mathrm {dis}}=\sqrt{A/\pi}$ is the radius of the bicelle of the membrane area, $A$.
The vesicle has bending energy of $8\pi(\kappa+\bar{\kappa}/2)$ and no edge energy.
An intermediate shape during the shape transition can be approximated as a spherical cap around the transition point \cite{nogu06a}.
Under this geometrical assumption, 
the excess energy, $\Delta E$, of the membrane with respect to the flat bicelle is given by \cite{Fromherz1983,Hu2012}
\begin{eqnarray}
 \label{eq:chp3eq1}
 &\Delta E (\Omega^{2},\zeta) = 4\pi(2\kappa + \bar{\kappa})\big[\Omega^{2} + \zeta(\sqrt{1 - \Omega^{2}} - 1)\big], \\
 \label{eq:chp3eq2}
 &\zeta = \dfrac{\Gamma R_{\rm {ves}}}{2 \kappa + \bar{\kappa}}, \:\: \Omega=\dfrac{R_{\rm {ves}}}{R}, \: \mbox{and} \: R_{\rm {ves}} = \sqrt{\dfrac{A}{4\pi}},
\end{eqnarray}
where $1/R$ is the curvature of the membrane.
The normalized curvature, $\Omega$, is an order parameter: $\Omega=0$ and $1$ for the bicelle and vesicle, respectively.
At $\zeta=1$, the bicelle and vesicle have the same energy and, for
$0<\zeta<2$, a free energy barrier exists at $0<\Omega<1$.

The free energy barrier can be determined by collecting samples in which
pre-curved membranes change into open disks or closed vesicles. 
The probability, $P(\Omega^{2})$, at which the pre-curved membranes
change into closed vesicles, is derived for the initial $\Omega$
as \cite{Hu2012} 
\begin{equation}
 P(\Omega^{2}) = \dfrac{ \int_{0}^{\Omega^{2}} dy \exp(\Delta
  E(y,\zeta)/\tilde{D})  }{ \int_{0}^{1} dy \exp(\Delta E(y,\zeta)/\tilde{D})  }.
\label{eq:pomega}
\end{equation}
By fitting this function to simulation data, $\tilde{D}$ and $\zeta$ are determined. 
The free energy barrier of eqn~(\ref{eq:chp3eq1}) is obtained from the value of $\zeta$.
We execute $512$ independent runs for each parameter set ($\Omega$, $N_{\mathrm{emb}}$), and calculate the free energy barrier between the bicelle and vesicle.
In the simulation, $N=41472$, $N_{\mathrm{amp}}=1000$, and $N_{\mathrm{emb}}=25$, 50,
100, 300, and 500 are used.

   \begin{figure}
    \centering
     \includegraphics[width=85mm]{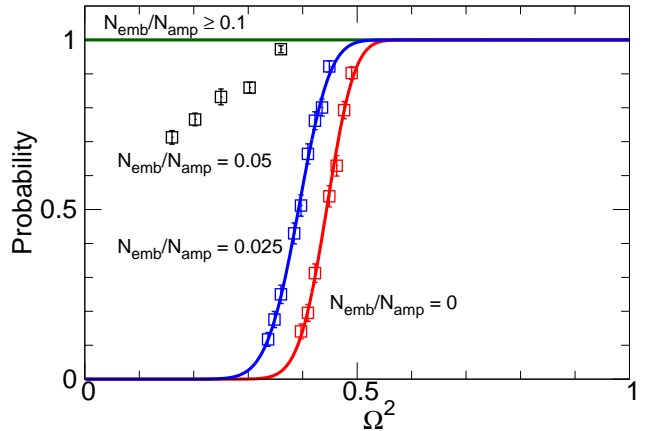}
    \caption{Transition probability to a vesicle as a function of $\Omega^{2}$.
    The squares represent the simulation data for $N_{\mathrm{emb}}/N_{\mathrm{amp}}=0$, $0.025$, and $0.05$. 
    The red and blue solid lines represent fits by eqn~(\ref{eq:pomega}) for $N_{\mathrm{emb}}/N_{\mathrm{amp}}=0$ and $0.025$, respectively.
    At $N_{\mathrm{emb}}/N_{\mathrm{amp}} \geq 0.1$, the
    transition probability becomes unity, independent of $\Omega^{2}$ (green solid line).}
    \label{fig:chp3six}
   \end{figure}

The $P(\Omega^{2})$ curves at $N_{\mathrm{emb}}/N_{\mathrm{amp}}=0$ and $0.025$
are well fit by eqn~(\ref{eq:pomega}) with $(\zeta,\tilde{D})=(1.49,0.00147)$ and $(1.56,0.00174)$, respectively (see Fig. \ref{fig:chp3six}).
As the hydrophobic
particles are increasingly embedded, $P(\Omega^{2})$ increases.
Thus, the free energy barrier is decreased by the embedded
hydrophobic particles. Surprisingly, no 
barrier exists for $N_{\mathrm{emb}}/N_{\mathrm{amp}}\geq 0.1$.
To clarify the causes of this decrease in the free energy barrier,
 we investigate the dependence of the membrane properties on  $N_{\mathrm{emb}}/N_{\mathrm{amp}}$ in the next two subsections.

\subsection{\label{chp3:sub1}Bending Rigidity}

We estimate the bending rigidity, $\kappa$, of the membrane containing hydrophobic
particles. 
Since the area of the tensionless membrane depends on
the fraction of embedded hydrophobic particles,
we set the tensionless state by changing  $N_{\mathrm{amp}}$ for a fixed value of $N_{\mathrm{emb}}$.
For each zero-surface-tension state, the method described in Section \ref{chp1:sub3:sub1} is applied to calculate $\kappa$.

   \begin{figure}
    \centering
     \includegraphics[width=70mm]{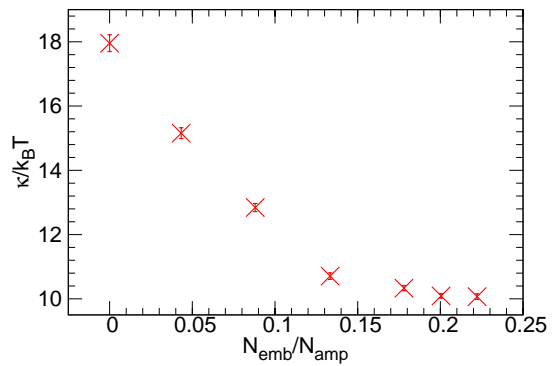}
    \caption{Bending rigidity, $\kappa$, dependence on $N_{\mathrm{emb}}/N_{\mathrm{amp}}$.}
    \label{fig:chp3two}
   \end{figure}

As the embedded hydrophobic particles are added, 
$\kappa$ first decreases 
and is saturated at $N_{\mathrm{emb}}/N_{\mathrm{amp}} \gtrsim 0.15$ (see Fig. \ref{fig:chp3two}). 
To explain this dependence,
we calculate two quantities of the bilayer membrane that may cause the change in the bending rigidity.
 One is  the bilayer thickness, $d$,
and the other is the orientational order parameter, $S$.
The orientational order is expressed by
\begin{equation}
 S = \langle\frac{3(\bm{n}_{i} \cdot \bm{n}_{z} )^2 - 1 }{2} \rangle,
\end{equation}
where $\bm{n}_{i}$ is the unit vector of the $i$-th amphiphilic molecule from
the hydrophobic to hydrophilic particles and $\bm{n}_{z}$ is the average normal direction of the membrane,
i.e., the unit vector along the $z$ axis.

For the homogeneous membrane in the absence of hydrophobic molecules,
$\kappa$ has the following relationships with
the orientational order and thickness. 
The orientational order of the amphiphiles is related to the membrane bending fluctuations
according to \cite{Watson2011,Watson2012}
\begin{equation}
 \langle |{n}^{||}_{\bm{q}}|^{2}\rangle = \frac{k_{\mathrm B}T}{\kappa q^{2}},
\end{equation}
where $n^{||}_{\bm{q}}=[\bm{q}\cdot {n}_{\bm{q}}]/q$,
${n}_{\bm{q}}$ is the Fourier transformation of the amphiphilic
orientational vector, ${\bm{n}}$. 
On the other hand, the membrane thickness, $d$, is related to $\kappa$ via the equation
\cite{landau1986course} 
\begin{equation}
 \label{eq:2}
 \kappa = \dfrac{1}{48}Yd^{3},
\end{equation}
where $Y$ is the Young modulus.
However, the heterogeneous membrane with embedded particles may not obey these relations.

   \begin{figure}[h]
    \centering
    \includegraphics[width=75mm]{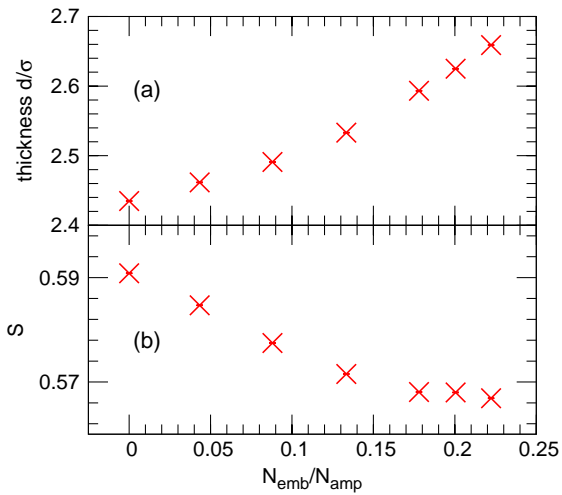}
    \caption{Dependence of (a) the orientational order parameter, $S$ and (b) the membrane thickness, $d$, on $N_{\mathrm{emb}}/N_{\mathrm{amp}}$.
    }
    \label{fig:chp3four}
   \end{figure}

The dependence of $d$ and $S$ of the embedded membranes on $N_{\mathrm{emb}}/N_{\mathrm{amp}}$ are shown in Fig. \ref{fig:chp3four}.
Following addition of the hydrophobic particles,
$d$ increases and $S$ decreases.
 The embedded hydrophobic particles disturb the
orientation of the amphiphiles, and enlarge $d$.
The decrease in $S$ is saturated at $N_{\mathrm{emb}}/N_{\mathrm{amp}} \gtrsim 0.15$ 
but $d$ shows no saturation.
   \begin{figure}[h]
    \centering
    \includegraphics[width=75mm]{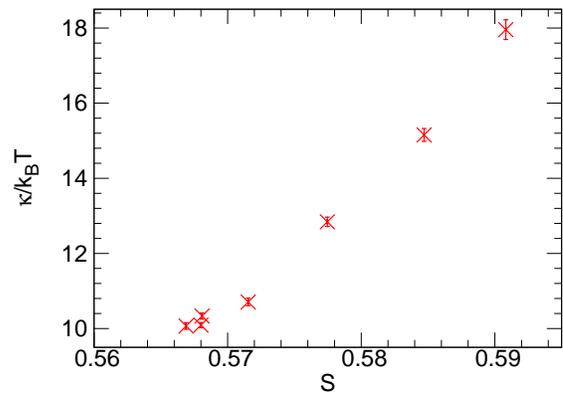}
    \caption{Bending rigidity, $\kappa$, as a function of the orientational order, $S$.}
    \label{fig:chp3thr}
   \end{figure}

According to eqn~(\ref{eq:2}), $\kappa$ increases with increasing $d$.
However, adding more hydrophobic particles decreases $\kappa$
despite the increasing $d$.
This is likely due to the rearrangement of the positions of 
 the hydrophobic particles in the bilayer.
In a curved membrane, the embedded hydrophobic particles can exist more in the outer (laterally expanded)
leaflet than in the inner leaflet.
In contrast, the change in
$S$ is similar to the change in $\kappa$. 
In order to illustrate this more clearly, we replot $\kappa$ as a function of $S$
in Fig. \ref{fig:chp3thr}.
The linear relation between $\kappa$ and $S$ is clear. 
Thus, the reduction of $\kappa$ by the embedded particles is mainly caused by
the disturbance of the orientation of the amphiphilic molecules.

\subsection{\label{chp3:sub2}Edge Line Tension}

\begin{figure}
    \centering
     \includegraphics[width=70mm]{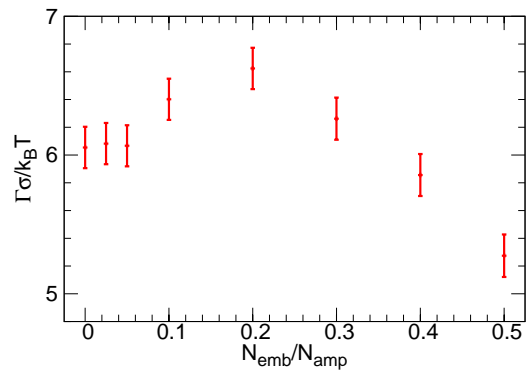}
    \caption{Edge line tension, $\Gamma$, as a function of $N_{\mathrm{emb}}/N_{\mathrm{amp}}$.}
    \label{fig:chp3fiv}
   \end{figure}

We estimate the edge line tension, $\Gamma$, at $N_{\mathrm{emb}}/N_{\mathrm{amp}}=0$ to 0.5
(see Fig. \ref{fig:chp3fiv}).
At $N_{\mathrm{emb}}/N_{\mathrm{amp}} \lesssim 0.1$,
$\Gamma$ is almost constant. 
For $N_{\mathrm{emb}}/N_{\mathrm{amp}} \in [0.1,0.3]$, 
$\Gamma$ increases, while for $N_{\mathrm{emb}}/N_{\mathrm{amp}} > 0.3$, 
$\Gamma$ decreases. 
At $N_{\mathrm{emb}}/N_{\mathrm{amp}} > 0.3$, 
the hydrophobic particles are concentrated at the membrane edge
and the membrane forms swollen round edges.
This reduction of the edge curvature would cause a
reduction in $\Gamma$.
For $N_{\mathrm{emb}}/N_{\mathrm{amp}} \in [0.1,0.3]$, the density of the
hydrophobic particles is not uniform on the membrane edge.
Thus, the membrane tends towards a shorter edge at which the hydrophobic particles can be uniformly distributed. 
This tendency likely enhances $\Gamma$.

\subsection{\label{chp3:sub4} Origin of Enhancement Effect}

By using the results shown in the previous two subsections, the
enhancement effect of the bicelle-to-vesicle transition is understood by the change in the membrane properties.
According to the results of Section \ref{chp3:sub2}, the change in
$\Gamma$ is apparent only for $N_{\mathrm{emb}}/N_{\mathrm{amp}} \geq 0.1$. 
However, the enhancement effect is apparent even for $N_{\mathrm{emb}}/N_{\mathrm{amp}} \simeq 0.025$, as shown in Fig. \ref{fig:chp3six}. Thus, the line tension is not a relevant parameter
for the enhancement.
In contrast, the change in $\kappa$ is apparent for
$N_{\mathrm{emb}}/N_{\mathrm{amp}} \lesssim 0.15$.

The ratio, $\bar{\kappa}/\kappa = -1.06$,
is obtained from the simulations at $N_{\mathrm{emb}}=0$.
This is a typical value for lipid membranes ($\bar{\kappa}/\kappa \simeq -1$) \cite{Hu2012}.
We assume that $\bar{\kappa}/\kappa$ is independent of $N_{\mathrm{emb}}/N_{\mathrm{amp}}$.
At $N_{\mathrm{emb}}/N_{\mathrm{amp}} \simeq 0.1$,
the  parameter, $\zeta \simeq 2.2$,
is obtained from $\kappa/k_{\mathrm B}T \simeq 12.4$.
This agrees with the results of the transition simulation, i.e., 
that all membranes transform into vesicles,
since the free energy barrier disappears at $\zeta \geq 2$.

In order to more quantitatively verify this,
 we compare two ratios $c_{\kappa}$ and $c_{\zeta}$, which are defined as 
\begin{equation}
 c_{\kappa} = \frac{\kappa(0)}{\kappa(0.025)} \ \mbox{ and } \
 c_{\zeta}  = \frac{\zeta(0.025)}{\zeta(0)},
\end{equation}
where $\kappa(0)$ and $\zeta(0)$ ($\kappa(0.025)$ and $\zeta(0.025)$)
denote $\kappa$ and $\zeta$ at $N_{\mathrm{emb}}/N_{\mathrm{amp}}=0$ ($0.025$), respectively.
When $\bar{\kappa} / \kappa$ and $\Gamma$ are constant, $c_{\kappa}=c_{\zeta}$.
In the region of $N_{\mathrm{emb}}/N_{\mathrm{amp}} \in [0,0.1]$,
$\kappa(N_{\mathrm{emb}}/N_{\mathrm{amp}})$ can be well fitted 
by the linear function, $\kappa(x)/k_{\mathrm B}T = ax + b$, for $(a,b)=(52 \pm 3, 17.5 \pm 0.3)$.
We calculate $\kappa(0.025)$ using this fit function. 
It gives $c_{\kappa}=1.07 \pm 0.03$ and $c_{\zeta}=1.045 \pm 0.002$, so that they coincide within the statistical error.
Thus, we conclude that the enhancement of the transition is caused by the reduction of $\kappa$.

The inclusion of molecules such as sterols and peptides modifies the bending rigidity, which also depends on lipid composition and buffer conditions \cite{dimo14,Gracia2010,fa07,yi12}.
The addition of cholesterols was previously believed to increase $\kappa$ in general \cite{deme10,dimo14}.
Recently, however, Dimova et al. reported that $\kappa$ is reduced for a sphingomyelin membrane and does not significantly change for DOPC membranes  with increasing cholesterol content \cite{dimo14,Gracia2010}. 
The reduction of $\kappa$ was also reported for the addition of the peptide, FP23, \cite{dimo14} and azithromycin \cite{fa07} to DOPC membranes.
The inclusion of lidocaine increases $\kappa$, whereas $d$ decreases in the L$_{\alpha}$ phase of DMPC/DMPG membranes \cite{yi12}.
Thus, depending on the embedded molecules, the bicelle-to-vesicle transition can be enhanced or suppressed.
In the present system, enhancement is obtained but the opposite effect (suppression) can be induced by a different type of inclusion. However, the effect can be predicted by measuring $\kappa$ and $\Gamma$ as functions of the inclusion density.

\section{\label{sum}Summary}

We have investigated shape evolutions of amphiphilic molecular assemblies
induced by the binding reaction between hydrophilic and hydrophobic molecules using DPD simulations. 
Various shape transformations 
such as bicelle-to-vesicle transition, multiple vesicle formation, vesicle fusion, 
and toroidal vesicle formation are exhibited.
During the shape changes, 
the hydrophobic particles embedded in the membrane
modify the membrane properties and affect the dynamics.
In particular, the reduction of the bending rigidity accelerates
the bicelle-to-vesicle transition.

The extension of discoidal arms from the vesicles and the closure of additional vesicles
are found in our simulations.
The formation of multiple vesicles from an oil droplet has been experimentally observed by Takakua et al. \cite{Takakura2003}. 
We expect that similar arm-extension and closure processes occurred in their experiment.

In this paper, we consider only a simple binding reaction.
The shape dynamics can be changed by reaction schemes and molecular compositions.
The effects of inverse and other reactions on the shape evolutions and its dependence on molecular architecture are important topics for further investigations.

\section*{Acknowledgments} 
We would like to thank T. Sugawara and T. Toyota for informative discussions. 
This work was supported by Japan Society for
the Promotion of Science through Program for Leading Graduate Schools
(MERIT) and a Grant-in-Aid for Scientific Research on Innovative Areas
'Fluctuation \& Structure' (No. 25103010) from the Ministry of
Education, Culture, Sports, Science, and Technology of Japan. The
numerical calculations were partly carried out on SGI Altix ICE 8400EX and FUJITSU PRIMEHPC FX10 at ISSP Supercomputer Center, University of Tokyo.

\footnotesize{

\providecommand*{\mcitethebibliography}{\thebibliography}
\csname @ifundefined\endcsname{endmcitethebibliography}
{\let\endmcitethebibliography\endthebibliography}{}

}

\end{document}